%% file: ms.tex
\documentclass[conference, a4paper]{IEEEtran}
\IEEEoverridecommandlockouts
\usepackage{cite}
\usepackage{amsmath,amssymb,amsfonts}
\usepackage{graphicx}
\usepackage{xcolor}
\usepackage{relsize}
\usepackage{hyperref}
\usepackage[hide]{todo}
\colorlet{todocolor}{red!70!black}
\def\todomark{\textcolor{todocolor}{To do}}
\usepackage{tikz}
\usetikzlibrary{positioning}

\begin{document}

  \title{Semi-supervised learning of glottal pulse positions in a neural analysis-synthesis framework}

  \author{\IEEEauthorblockN{Frederik Bous}
      \IEEEauthorblockA{\textit{UMR 9912 STMS} \\
      \textit{IRCAM, Sorbonne University, CNRS}\\
      Paris, France \\
      frederik.bous@ircam.fr}
    \and
      \IEEEauthorblockN{Luc Ardaillon}
      \IEEEauthorblockA{\textit{UMR 9912 STMS} \\
      \textit{IRCAM, Sorbonne University, CNRS}\\
      Paris, France \\
      luc.ardaillon@ircam.fr}
    \and
      \IEEEauthorblockN{Axel Roebel}
      \IEEEauthorblockA{\textit{UMR 9912 STMS} \\
      \textit{IRCAM, Sorbonne University, CNRS}\\
      Paris, France \\
      axel.roebel@ircam.fr}
    \thanks{This work has been funded partly by the ANR project ARS (ANR-19-CE38-0001-01)}
  }

  \maketitle

  \begin{abstract}
    \input{abstract.tex}
  \end{abstract}


  \input{paper.tex}

  \done\todo{Bous: Remove dublicates from refs\_luc.bib}
  \bibliographystyle{IEEEbib/IEEEtran}
  \bibliography{IEEEbib/IEEEabrv,refs}

\todos
\end{document}

%% file: abstract.tex
\done\todo{Axel: Fix abstract}
This article investigates into recently emerging  approaches that use deep neural networks
for  the estimation  of  glottal closure  instants (GCI).
We build  upon  our previous approach
that used synthetic speech exclusively
to create  perfectly annotated training data
and that had
been shown to compare favourably with other training approaches using 
electroglottograph (EGG) signals.  Here we
introduce a semi-supervised training strategy that allows refining  the estimator
by means of  an analysis-synthesis setup using real speech signals,
for which  GCI ground truth does  not exist.  Evaluation  of the analyser is  performed by
means of comparing the GCI extracted from the glottal flow signal generated
by the  analyser with the  GCI extracted from EGG on the  CMU arctic
dataset, where EGG signals were recorded in addition to speech.  We observe that (1.)  the
artificial increase of  the diversity of pulse  shapes that has been used  in our previous
construction of  the synthetic database is  beneficial, (2.)  training the  GCI network in
the analysis-synthesis setup allows achieving a  very significant improvement of the GCI
analyser, (3.)   additional regularisation strategies  allow improving the  final analysis
network when trained in the analysis-synthesis setup.


%% file: paper.tex
\newcommand{\rd}{\ensuremath{R_d}}
\newcommand{\ie}{i.\,e.}
\newcommand{\eg}{e.\,g.}
\newcommand{\fo}{\ensuremath{f_0}}
\newcommand{\sota}{state-of-the-art}
\newcommand{\textrightarrow}{$\,\to\,$}
\newcommand{\unit}[1]{\ensuremath{\mathrm{#1}}}
\newcommand{\configuration}[1]{\textsf{\textsmaller[0.5]{#1}}}
\newcommand{\fcna}{\configuration{FCN-Agamemnon}}
\newcommand{\fcnb}{\configuration{FCN-baseline}}
\newcommand{\voca}{\configuration{Anasynth-A}}
\newcommand{\vocb}{\configuration{Anasynth-B}}
\newcommand{\refspec}{\configuration{SR}}
\newcommand{\sedreams}{\configuration{SEDREAMS}}
\newcommand{\loss}[1]{\textsf{\textit{\textsmaller[0.5]{#1}}}}
\newcommand{\ASs}{\loss{AS-spectral}}
\newcommand{\ASt}{\loss{AS-time}}
\newcommand{\ASAt}{\loss{ASA-time}}
\newcommand{\As}{\loss{A-spectral}}
\newcommand{\At}{\loss{A-time}}

\section{Introduction}
Glottal Closure Instants (GCI) detection
consists in finding the temporal locations
of significant excitation of the vocal tract
that occur in voiced speech
during the closure of the vocal folds.
GCI detection has many applications
in speech analysis and processing \cite{drugman2014excitation},
among which analysis of vocal disorders \cite{reddy2018glottal, deshpande2018effective},
formants estimation \cite{anand2006extracting},
or speech synthesis \cite{drugman2012deterministic}.

One possible way to extract the GCI positions
in a speech signal
is to use parallel electroglottographic (EGG) recordings
using dedicated hardware
that measures the vocal folds contact area.
GCIs can then be extracted using peak-picking on the derivative of the EGG signal.
However, such recordings are rarely available,
motivating the development of methods
that can extract GCIs directly from speech signals.
Many algorithms
have been proposed for this purpose.
Until recently,
most approaches
used to be based on hand-crafted
digital signal processing techniques and heuristics.
Typically, such methods
first compute an intermediate speech representation,
such as the linear prediction residual
\cite{naylor2007estimation, prathosh2013epoch},
a zero-frequency filtered signal \cite{murty2008epoch},
or a smoothed signal \cite{drugman2009glottal},
which emphasises the locations
of glottal closure instants
found at local maxima,
impulses or discontinuities.
Then, dynamic-programming or peak-picking
is used to select the GCIs among the detected candidates. 

Although such approaches have been shown to perform reasonably well,
they sometimes rely on manual parameter tuning
(\eg, the mean \fo\ for SEDREAMS)
and the quality of their results
remains quite dependant
on the characteristics of the analysed speech signal
(\eg, pitch and voice quality, speech or singing voice) \cite{babacan2013quantitative}.
Moreover, some algorithms like SEDREAMS \cite{drugman2009glottal}
or DYPSA \cite{naylor2007estimation}
also detect GCIs during unvoiced segments
and thus rely on further algorithms
to filter out GCI candidates in unvoiced parts. 

To overcome the limits of those methods,
new data-driven approaches have been recently proposed.
In \cite{yang2018detection}, authors used a classification approach
where GCI candidates are the negative peaks
of a low-pass filtered signal.
Similarly, \cite{reddy2018glottal} also employed a classification-based approach
using 3 parallel CNNs
operating on different signal representations.
In \cite{goyal2019detection},
the authors used a CNN
to optimise both a classification and regression cost,
where a GCI is simultaneously detected
and localised in a frame.
In other related works \cite{prathosh2019adversarial, deepak2019glottal},
the authors used neural networks
to perform a regression from the speech waveform
to the corresponding EGG signal
using 
adversarial training procedures. 

All these approaches rely on EGG signals
for training the networks,
which has two main drawbacks:
(1.) the EGG signals are often noisy,
and the extracted ground-truth GCIs
are thus likely to contain errors;
(2.) EGG signals are rarely available,
which makes it difficult
to build large multi-speaker databases for training,
and thus limits the generalisation ability of the models. 

Recently, we proposed to use high quality synthetic speech signals with almost perfect
annotation for training a neural network to predict the glottal flow signal from raw
speech \cite{ardaillon2020gci}. In these initial experiments we have shown that despite
being trained on synthetic speech, the resulting network compares favourably with other
methods trained on EGG data when evaluated on real speech.
While training with synthetic signals allows creating nearly unlimited training data with
reliable annotation covering many different pulse shapes, it will not cover the full
variability (pulse forms, jitter, shimmer) encountered in real speech. Notably, rough or
creaky voices with irregular pulses are difficult to synthesise realistically.

\done\todo{Axel: reviewers need a short summary of the novelties, describe these at the end of the introduction}
To this end the following paper introduces a new semi-supervised approach that allows
extending the synthetic dataset with real speech signals that do not
have any associated EGG signal or GCI annotation.
This is achieved by means of a complete analysis-synthesis framework that
uses the detected pulses for resynthesising the original speech signal. Our evaluation
shows that this analysis-synthesis setup allows significantly improving the performance
on real speech.

In Section \ref{sec:approach}
we will present an overview of
the proposed approach,
network architecture,
databases,
and training procedure.
The methodology
and results of our evaluations
will be presented in Section \ref{sec:evaluation}.
Finally, we will draw a conclusion
in Section \ref{sec:conclusion}
\done\todo{Bous: Check before final version}

\section{Proposed Approach}
\label{sec:approach}

\subsection{The Synthetic Data Base}
The power of deep learning depends on the available training data. While there exist
numerous speech databases with speech, only few of these databases also contain EGG
signals. Therefore GCI extraction methods that rely on the presence of EGG signals cannot
fully exploit the benefits of deep learning based methods.
 
In previous work \cite{ardaillon2020gci}, we proposed to use a synthetic database in order
to have glottal flow signals available as ground truth. The dataset consists of the
publicly-available BREF \cite{gauvain1990design} and TIMIT \cite{zue1990speech} datasets, that were
analysed   and    re-synthesised   using    the   PaN    vocoder   \cite{huber2015glottal,
  ardaillon2017synthesis}.    The    PaN   vocoder   uses the LF    glottal   source   model
\cite{fant1985four, fant1995lf} as a  source signal. For the resynthesis the \fo\
estimation of \cite{ardaillon2019fully} was used to generate the source signal by creating
pulses which are separated by the fundamental period. Thus an almost perfect ground truth
glottal pulse can be generated from the vocoder model.

By means of gathering the speech recordings available from previous projects we increased
the size of the dataset from a few hours to about 165 hours of human voice including
normal   speech, singing,   whispering   and   shouting.   We   call   our new   dataset
\emph{Agamemnon}. The simplified LF-model described in \cite{fant1995lf}, which is used
by the PaN vocoder, uses a single parameter to describe the pulse, the \rd\ parameter. In
\cite{ardaillon2020gci} we resynthesised the database multiple times using different
\rd-values in order to create a large variety of different pulse shapes. Assuming that
the increased database size would provide a sufficiently large variety of \rd\ values,
only one resynthesis of the database using the \rd-values provided by the \rd\ extraction
algorithm \cite{degottex2010phase, huber2012glottal} was used in this work.

The training database thus contains a resynthesis of all speech signals produced by the
PaN synthesiser annotated with all PaN parameters (GCI, \rd, \fo, voiced/unvoiced segments,
spectral envelops and unvoiced components (see below)), and all real speech signals for
which a restricted set of annotations will be used (\fo, voiced/unvoiced segments, \rd\ analysis,
spectral envelops).

\subsection{The Analysis-Synthesis Framework}
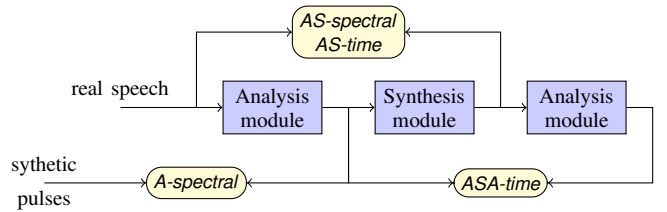
\begin{figure}[t]
  \centering
  {\input{fig_vocoder.tex}
   \done\todo{Bous: it seems to me that the term \emph{spectral} in the ana spectral loss
      should disappear because there is also an ana time loss that otherwise does not fit
      into the schema. }}
  \caption{
    Schematic of the analysis-synthesis framework and
    a visualisation of Table \ref{tab:vocodersetup}.
    Neural networks are in blue,
    losses are in yellow.
    The synthesiser has as additional input
    a noise source signal
    and the spectral envelopes of the audio
    and of the unvoiced signal component.
    The two analysis modules share the same weights.
    The \At\ loss is calculated separately
    on the synthetic speech
    (not included in this figure).
  }
  \label{fig:vocoder}
\end{figure}
\begin{table*}[t]
  \centering
  \caption{
    The loss function of the analysis-synthesis framework is split into five separate losses
    that are all calculated on different ``outputs'' of the framework,
    each with its own target value.
    The details for each component are given below.
  }
  \label{tab:vocodersetup}
  {\input{tab_losses}}
  \done\todo{Bous: the loss terms are very confusing, you should try to better distinguish
    the output (e.g. Ana-synth) and the type of loss (time or spectral). Currently it is
    very difficult to distinguish these two parts. This comes in part due to the fact
    that spectral and time losses are not explicitly introduce in the text. You could
    discuss those in the section losses. sIt is also not clear how one should
    read the output column? It says output but in fact describes the input and
    processing path. Can you make this easier to grasp?}
\end{table*}

To allow the glottal flow extraction module
to model the various phenomena from real speech 
that are not captured by the PaN resynthesis,
we train it
in an analysis-synthesis framework.
In this framework,
an analyser (the glottal flow extraction module)
and a synthesiser (see below)
are cascaded and trained simultaneously.

Following the approach from \cite{ardaillon2020gci}
the modules of this paper also work on audio
given at a sample rate of $16\unit{kHz}$.
Due to pooling layers in the analyser,
its output sample rate is $2\unit{kHz}$.
To match the required input sample rate
of the synthesiser,
upsampling using linear interpolation
is performed during training.
When inferring the GCI,
upsampling is performed
using cubic splines
in order to increase accuracy.

\paragraph*{Synthesiser}
We use a convolutional neural network
with dilated convolution based on WaveNet
\cite{oord2016wavenet}
to transform the glottal pulses
generated by the analysis module,
and an additional white-noise source,
to raw audio.
\done\todo{Axel: Add citations and explanations}
Additionally, we provide as control parameters,
the spectral envelopes of the speech signal,
represented in 64 mel-bands,
as well as the spectral envelopes
of the unvoiced speech component,
represented in 16 mel-bands.
The unvoiced component is extracted
following the remixing approach described in \cite{huber2015glottal};
the spectral envelopes are calculated according to \cite{roebel2007cepstral}.
\done\todo{Bous: what parameters}

Unlike the original WaveNet,
and similar to \cite{wang2019neural}
and parallel WaveNet \cite{oord2017parallel}
we create whole frames in one forward pass.
Avoiding the recursive structure of the original WaveNet is possible
due to the different objective:
WaveNet requires recursion to create periodic signals.
Our synthesiser has the periodic pulses as input
and thus performs not much more
than filtering the input
according to the spectral envelopes.
Furthermore, since the pulses
are coherent with the speech signal,
the ambiguity in the phase is minimised
for the voiced speech component.
Consequently, rather than learning
a general distribution of raw audio speech signals
under their control parameters,
here we can use the speech signals as \emph{ground truth}
and require that the actual audio at hand is properly reproduced
according to appropriate loss metrics (see below),
which sufficiently models the speech properties
when the objective is to infer the pulse positions.

Since the synthesis task is much simpler
than for the original WaveNet,
even simpler than in \cite{shen2017natural},
we follow the observation from \cite{shen2017natural}
and use a reduced size with only
two stacks of six layers each
(rather than three stacks of ten layers each).

\subsection{Loss functions}

The following section describes the different loss functions that are used to train the
networks. Figure \ref{fig:vocoder} gives a schematic overview
of the various losses used to update the networks
and the flow of the signals used to calculate them.

To train the analysis-synthesis loop, we follow \cite{wang2019neural}, using
multi-resolution spectral loss for the resynthesis, \ie, the mean absolute error (MAE)
between multiple STFT
with different window lengths, which we obtain by multiplying the output segment
\done\todo{Bous:Clarify whether this is a magnitude only loss}%
with Hann windows of given sizes,
(see Table \ref{tab:vocodersetup}),
applying an FFT to each of the windowed segments
and taking the logarithm of the absolute values.
The overlap is $1/2$ for each resolution.
We call this loss \ASs.

The STFT considers only absolute values of each frequency bin.
Phase differences, therefore, do not affect the spectral loss.
This property is required to properly model
the noise component of the signal
where the phases are random.
On the other hand, the pulse positions
are encoded in the phases as well.
The spectral loss is thus not suitable
to enforce proper pulse positions.

Since the analysis module aims to provide a coherent representation for the
excitation signal of both the real and the resynthesised speech signal, there is no
ambiguity in phase. We can thus also apply a loss in the time domain between the real
audio and the resynthesis to enforce proper reconstruction of the phases,
which shall be referred to as \ASt.
\done\todo{Bous: looks as if you only calculate the loss here, is there not also an update
  involved? Please clarify.
  Which loss is applied where is clarified under Training procedure.%
}
The time domain loss however will not correctly deal with
noise. Therefore, when calculating the loss in the time domain, the noise input of the
synthesiser is set to zero, enforcing the synthesiser to only produce the deterministic
waveform.

When training the GCI analyser in the analysis-synthesis setup,
the analyser and synthesiser networks
need to be constrained
such that they focus on their respective tasks.
Notably, the analyser should not start to represent more than the glottal pulse
form in its output even if such a strategy clearly would help to improve the synthesis.
For imposing these constraints we rely on two strategies:
In each training step,
the analyser is also trained on the synthetic data,
just like in \cite{ardaillon2020gci}, creating the \At\ loss,
and additionally, we include the following regularisation losses.



\paragraph{\ASAt}

In the analysis-synthesis setup
it is desired
that the original signal
and the resynthesised signal
have the same glottal pulses.
Thus we reanalyse the resynthesised audio
and apply a loss in the time domain between
the reanalysis
and the original analysis.

Note that a global optimum for this loss is reached
if the analyser does not produce any pulses.
In that case all pulse signals are zero
and as a result the loss is trivially zero.
This behaviour was observed
when the synthesiser was not pre-trained
and no other regularisation was performed
on the analysis of the real audio.

\paragraph{\As}
While we do not know the exact pulse positions
in the real audio,
we know where pulses are present
(the voiced segments),
the approximate pulse shape (from the \rd-analysis),
and the pulse frequency (from the \fo-analysis).
All these parameters are also
in the synthetic glottal pulse signal.
Therefore the spectrograms should be similar
for the two variants of each speech signal, that are
the synthetic and the real version.
To enforce this invariant,
we apply a spectral loss between
the synthetic glottal pulse signal
and the analysed glottal pulse signal.
This loss is considered a major regularisation of the analyser
avoiding the network to create two distinct behaviours for
real and synthetic speech.

\subsection{Training procedure}
\label{ssec:procedure}
\subsubsection*{Step 1 -- Initialisation}
Both the analysis and the synthesis modules
are trained separately on the synthetic database.
For the analysis module we use the same hyper-parameters
as in \cite{ardaillon2020gci}, that is,
a batch size of $128$,
an epoch length of $500$ updates,
an initial learning rate of $2\cdot10 ^ {-4}$ and
decrease the learning rate by a factor of $0.75$
if no improvement on the validation loss
was observed for 10 epochs.
The synthesiser is trained with
a batch size of $8$ training-samples
from $8$ different files and a length of $2560$ audio-samples,
an epoch length of $512$ updates,
an initial learning rate of $5\cdot10 ^ {-5}$
and the same learning rate schedule as for the analyser
with a minimum of $1\cdot10^{-6}$.\done\todo{Bous: moved from second step, this relates to the first step no ?}
After pre-training,
the batch-normalisation layers in the analyser are frozen.

\subsubsection*{Step 2 -- Analysis-Synthesis}
The analysis and synthesis modules
are combined in the analysis-synthesis setup.
Both real and synthetic data are used
to generate the different losses.
For details see Table \ref{tab:vocodersetup} and
Figure \ref{fig:vocoder}.
Each training step consists of two updates.
First the losses involving real data
 (\ASs,
  \ASt,
  \ASAt,
  \As)
are calculated according to Table \ref{tab:vocodersetup}
and the gradients are propagated to both analyser and synthesiser.
\done\todo{%
  Bous: the second analyser at the output should not be adapted
  with the \ASAt! Is this the case, please clarify.%
}
Then a ``conventional'' training step
(as in Step I)
is performed for the analyser
on synthetic data
to ensure that the output of the analyser
does not diverge too far from the LF-model.

For the analysis-synthesis we use
a batch size of $8$
with a length of $3553$ audio samples
($2560$ remain as input to the synthesiser
 due to ``valid''-padding in the analyser),
an initial learning rate of $1\cdot10^{-5}$
with the usual schedule (cf.\ Step 1)
with no minimal learning rate.

\section{Evaluation}
\label{sec:evaluation}
\begin{table}
  \caption{Results on CMU with metrics considering all GCI}
  \label{tab:results}
  \input{tab_results}
\end{table}

We evaluate the performance of the analyser
on the CMU arctic \cite{kominek2004cmu} databases
using the metrics from \cite{ardaillon2020gci},
which are based on the metrics defined in \cite{thomas2009sigma}.
The results are summarised in Table \ref{tab:results}.
The GCI are obtained
using a simple peak-picking procedure
on the negative peaks
of the derivative of the predicted glottal flow,
as was done in \cite{ardaillon2020gci}.

\subsection{Test Setup}
We compare the following configurations:

\begin{itemize}
  \item{\textbf{\sedreams}}
    For comparison,
    we include the results from
    SEDREAMS \cite{drugman2009glottal}
    as a baseline.

  \item{\textbf{\fcnb}}
    As another baseline
    we use the configuration
    ``FCN-synth-GF'' from \cite{ardaillon2020gci}.
    This is assumed to be the current \sota.

  \item{\textbf{\fcna}}
    The same network and training procedure
    as \fcnb\
    but trained on the synthetic (not augmented) part of the
    \emph{new}, larger database (Agamemnon). 

  \item{\textbf{\voca}}
    The analysis module from the analysis-synthesis
    as introduced in the previous section.
    It was initialised with \fcna.

  \item{\textbf{\vocb}}
    The analysis module from the analysis-synthesis,
    however trained without
    \As.
    Also initialised with \fcna.

    \done\todo{Bous: Ana time loss has never been introduced, if I
      see correctly this is he loss from \cite{ardaillon2020gci}?}    
\end{itemize}

\subsection{Evaluation Metrics}
We use the same evaluation metrics
that were used for Table 2 in \cite{ardaillon2020gci},
which are defined in Table \ref{tab:metrics}.
For each model
we calculate the metrics for each file separately
and create an average for each of the speakers,
\emph{bdl}, \emph{jmk} and \emph{slt}.
We then create total averages,
as the average over each speaker,
\ie, equally representing each \emph{speaker},
and display the result in Table \ref{tab:results}.


\begin{table}
  \caption{Evaluation metrics used in Table \ref{tab:results}}
  \label{tab:metrics}
  \begin{tabular}{lp{7.5cm}}
    IDR & (\emph{identification rate})
      The number of unique identifications
      between generated pulses
      and pulses in the ground truth
      divided by the total number of pulses in the ground truth. \\
    MR & (\emph{miss rate})
      The number of pulses in the ground truth
      that were not detected by the system
      divided by the total number of pulses in the ground truth. \\
    FAR & (\emph{false alarm rate})
      The number of generated pulses
      that could not be associated with any pulse
      from the ground truth
      divided by the total number of pulses in the ground truth. \\
    IDA & (\emph{identification accuracy})
      Standard deviation of the misalignment
      between the associated detected and ground truth GCI. 
  \end{tabular}
\end{table}

\subsection{Observations}

\subsubsection{Having a large variety of pulse shapes
               in the synthetic data is important}
Although the database for \fcna\ was much larger
than the database for \fcnb,
the identification rate of \fcna\
is slightly worse than the one of \fcnb.
This can be explained by the larger variety
of \rd\ values in the synthetic database for \fcnb\ 
obtained by the data augmentation
(multiple resynthesis with different \rd-values).

It is worth noting that
although this paper focuses on estimating
the glottal closure instances (GCI),
the analyser learns much more than just the GCI:
It learns the whole pulse shape, that is,
in the context of the underlying pulse model
(LF-model \cite{fant1995lf})
it learns the proper \rd-value.
Thus augmenting the variety in \rd-values/pulse-shapes
improves generalisation.
Furthermore,
the pulses generated by the analysis module
could potentially be used to infer the \rd-value,
which might improve the \rd-analysis
and consequently the resynthesis of the synthetic database.

\subsubsection{Refining the analysis as part of the analysis-synthesis framework
               does improve the GCI detection}
We see that the identification rate of \voca\
surpasses its initialisation \fcna\
and both baselines.
We can conclude that refining the analysis
in the analysis-synthesis framework
does positively impact the performance of the analysis module.

This can be explained by the fact
that the refinement happens on real data
and the real data is closer to the evaluation dataset,
CMU arctic,
than the synthetic data
and suggests that
the analysis-synthesis setup is able to
force the analysis module
to generate an expressive pulse model.
Consequently the model generalises better to real data,

Observe that none of the losses used for training
explicitly represents the GCI.
Nevertheless our model is able
to detect the GCI better than any other method.
This suggests that the model
does not only learn the pulse positions,
it really learns the pulse shapes
from which the GCI can be derived.

\subsubsection{Regularisation of the pulse signal is crucial}
The improvement of \vocb\ is smaller than for \voca\ compared to \fcna\
and still does not allow to surpass the results of \fcnb.
This shows that 
the regularisation in terms of the \As\
is a crucial element of this set-up.
One problem with the analysis-synthesis method
is the fact
that the synthesiser learns
only to use the pulse input
in voiced speech.
Consequently the predicted pulse signal
has no effect at the synthesiser output
during unvoiced speech and silence
and thus does not affect the \loss{AS} losses.
Since most of the error
happens in the unvoiced segments,
the improvement in configuration \vocb\ is minimal.
\done\todo{Bous: I don't follow you here. Your argument explains why the false alarm rate is of
Anasynth-NB is higher than Anasynth-A. Not why the IDR is better for Anasynth-A. Am I
wrong?}

\done\todo{%
  Bous: unclear? Why would it not generate any pulses ?
  I feel the experiment with SR case is not well explained. May be you should remove it and
  just summarise the result in a single phrase: In an additional experiment it was shown that
  Training real data with the Ana spectral loss for real data and the Ana time loss for
  synthetic
  data does not allow improving over the FCN Agamemnon baseline.%
}

In an additional experiment it was shown that
training the analyser on real data with \As\
and on synthetic data with \At,
hence avoiding the use of a synthesiser,
does not allow improving over the \fcna\ baseline.

\section{Conclusion}
\label{sec:conclusion}
In order to further improve our previous results
\cite{ardaillon2020gci},
we proposed a new semi-supervised approach
for training a neural GCI detector
that allows training
with controlled synthetic speech
and real speech data.
The main idea of this approach
is to incorporate our previously-proposed neural network
for predicting the glottal flow
into a constrained analysis-synthesis loop
trained using a reconstruction loss
and a collection of additional loss functions that ensure proper
segregation of the analysis and synthesis tasks.
To better represent all the variability of voice signals,
we also proposed to use here
a much larger database
with many different types of signals.

For the task of GCI detection,
evaluation on the CMU arctic database has shown
that the refinement step using real data
significantly improves
the identification rate of the estimator
with respect to training it solely
on the synthetic database.
The identification rate also surpasses
all previous systems
establishing the proposed method
as the new \sota.
While we only assessed in this paper
the performance of our analyser for GCI detection,
the predicted glottal flow may also be used
for other tasks like \rd\ analysis.


%% file: fig_vocoder.tex
\tikzstyle{model} = [draw, fill=blue!20,   minimum width=1.3cm, align=center]
\tikzstyle{loss} =  [draw, fill=yellow!20, align=center, rounded corners=2mm]

\footnotesize
\newcommand{\size}{1.0cm}
\newcommand{\losss}[1]{\loss{\scriptsize #1}}
\begin{tikzpicture}[node distance=\size]
  \node[model] (synth) {Synthesis \\ module};
  \coordinate[left of=synth] (as) {};
  \coordinate[right of=synth] (sa) {};
  \node[model, left of=as] (ana1) {Analysis \\ module};
  \coordinate[left of=ana1] (in) {};
  \coordinate[left of=in] (realin) {};
  \node[model, right of=sa] (ana2) {Analysis \\ module};

  \node[loss, above of=as] (loss1) {\ASs \\ \ASt};
  \node[loss, below of=sa] (loss2) {\ASAt};
  \node[loss, below of=in] (loss3) {\As};
  \coordinate[left of=loss3] (synthin0) {};
  \coordinate[left of=synthin0] (synthin) {};

  \node[above=0pt of realin] {real speech};
  \node[above=0pt of synthin] {sythetic};
  \node[below=0pt of synthin] {pulses};

  \draw[->] (realin) -- (ana1) ;
  \draw[->] (ana1) -- (synth);
  \draw[->] (synth) -- (ana2);
  \draw[->] (in) -- ++(0,\size) -- (loss1);
  \draw[->] (sa) -- ++(0,\size) -- (loss1);
  \draw[->] (ana2) -- ++(\size,0) -- ++(0,-\size) -- (loss2);
  \draw[->] (as) -- ++(0,-\size) -- (loss2);
  \draw[->] (as) -- ++(0,-\size) -- (loss3);
  \draw[->] (synthin) -- (loss3);

\end{tikzpicture}

%% file: tab_losses.tex
\begin{tabular}{|l|l|l|l|l|l|}
  \hline
  loss name & input 1 (network output) & input 2 (target reference) & weight  & loss function & window sizes in \unit{ms}\\
  \hline
  \ASs      & real audio 
              \textrightarrow\ ana 
              \textrightarrow\ synth   & real audio       & $0.1$   & MAE  & $160$,
                                                                            $53.3$,
                                                                            $32.0$,
                                                                            $22.9$,
                                                                            $17.8$\\
  \ASt      & real audio 
              \textrightarrow\ ana 
              \textrightarrow\ synth 
              ($\text{noise}=0$)      
                                       & real audio       & $10$    & MAE  & point wise (time-domain)\\
  \ASAt     & real audio 
              \textrightarrow\ ana 
              \textrightarrow\ synth
              \textrightarrow\ ana     & real audio 
                                         \textrightarrow\ ana 
                                                          & $1$    & MSE  & point wise (time-domain)\\
  \As       & real audio 
              \textrightarrow\ ana     & synthetic (PaN) pulses 
                                                          & $0.02$ & MAE  & $80$, $40$\\
  \At       & synthetic (PaN) audio
              \textrightarrow\ ana     & synthetic (PaN) pulses 
                                                          & $10$   & MSE  & point wise (time-domain)\\
  \hline
\end{tabular}

%% file: tab_results.tex
\centering
\begin{tabular}{|lrrrr|}
  \hline  model & IDR & MR & FAR & IDA\\
   & (in \%) & (in \%) & (in \%) & (in \unit{ms})\\
  \hline
  \voca & $94.21$ & $1.66$ & $9.71$ & $0.51$\\
  \vocb & $92.55$ & $0.87$ & $15.21$ & $0.64$\\
  \fcna & $91.90$ & $0.88$ & $22.33$ & $0.75$\\
  \fcnb \cite{ardaillon2020gci} & $92.73$ & $2.86$ & $10.87$ & $0.47$\\
  \sedreams \cite{drugman2009glottal} & $90.74$ & $0.19$ & $61.38$ & $0.30$\\
  \hline
\end{tabular}